\documentclass[conference]{IEEEtran}
\IEEEoverridecommandlockouts
\usepackage{cite}
\usepackage{amsmath,amssymb,amsfonts}
\usepackage{algorithmic}
\usepackage{graphicx}
\usepackage{textcomp}
\usepackage{xcolor}
\usepackage{algorithm}
\usepackage{algorithmic}
\usepackage{caption}

\def\BibTeX{{\rm B\kern-.05em{\sc i\kern-.025em b}\kern-.08em
    T\kern-.1667em\lower.7ex\hbox{E}\kern-.125emX}}
\begin{document}

\title{Patient-Specific Seizure Prediction Using Single Seizure Electroencephalography Recording
}

%
\author{\IEEEauthorblockN{Zaid Bin Tariq\IEEEauthorrefmark{1},
		Arun Iyengar\IEEEauthorrefmark{2},
		Lara Marcuse\IEEEauthorrefmark{3},
		Hui Su\IEEEauthorrefmark{1} and
		B\"{u}lent Yener\IEEEauthorrefmark{1}}
	\IEEEauthorblockA{\IEEEauthorrefmark{1,2}
		Rensselaer Polytechnic Institute,
		Troy, NY, USA \\}
	\IEEEauthorblockA{\IEEEauthorrefmark{2}
		IBM Thomas J. Watson Research Center, Yorktown Heights, NY, USA}
	\IEEEauthorblockA{\IEEEauthorrefmark{3}
		Mount Sinai Hospital
		New York, NY, USA}}
\maketitle

\begin{abstract}

Electroencephalogram (EEG) is a prominent way to measure the brain activity for studying epilepsy, thereby helping in predicting seizures. Seizure prediction is an active research area with many deep learning based approaches dominating the recent literature for solving this problem. But these models require a considerable number of patient-specific seizures to be recorded for extracting the preictal and interictal EEG data for training a classifier. The increase in sensitivity and specificity for seizure prediction using the machine learning models is noteworthy. However, the need for a significant number of patient-specific seizures and periodic retraining of the model because of non-stationary EEG creates difficulties for designing practical device for a patient. To mitigate this process, we propose a Siamese neural network based seizure prediction method that takes a wavelet transformed EEG tensor as an input with convolutional neural network (CNN) as the base network for detecting change-points in EEG. Compared to the solutions in the literature, which utilize days of EEG recordings, our method only needs one seizure for training which translates to less than ten minutes of preictal and interictal data while still getting comparable results to models which utilize multiple seizures for seizure prediction.
\end{abstract}


\section{Introduction}
\label{Into}
Epilepsy is a neurological disorder with 0.5\%-1\% of the world population suffering from the disease. It involves a recurrent occurrence of seizures \cite{fisher2005epileptic}. The seizures are characterized by sudden bursts of electrical activity from neurons of a focus region of the seizure \cite{valentinuzzi2007bioelectrical}. This abnormal firing of neurons results in a synchronous activity in the focus region \cite{blume2001glossary}. These seizures can be classified as primary generalized epilepsy which results in widespread electrical discharge which involves most of the brain \cite{van2014functional} or partial seizures, which can be classified by studying the  brain lobe involved, like temporal, frontal, occipital and parietal lobe
epilepsy \cite{van2014functional}.  For the patient with epilepsy, occurrences of seizures are unpredictable and may result in unconsciousness, or impaired tonic and clonic movement \cite{van2014functional}. This unpredictability has many adverse psychological and social effects on the daily life of the patient in addition to being life-threatening \cite{daoud2019efficient}. 

In order to reduce these adverse effects, anti-epileptic drugs (AEDs) are prescribed by doctors for suppressing seizures. The AEDs are successful for 60\%-70\% of the epilepsy patients, and for 40\%-60\% of these patients, the use of AEDs can be withdrawn after a considerable time without recurrence of seizures \cite{brodie2002staged}. For the patients for whom the AEDs are not beneficial, one treatment involves the resection of epileptic focus region. The focus region is estimated by localizing the origin of seizure using Electroencephalography (EEG). The rhythmical ictal  discharges are reflected in the EEG with higher amplitude during the seizures periods (ictal region) compared to the normal state (interictal region) of the patient. Using extracranial and intracranial EEG, these epileptic discharges serve as a way of pinpointing the focus region of the seizure.  

A system for predicting seizures can greatly help in improving the life of the patient with epilepsy. The system can act as as a useful indicator for warning the stakeholders involved about the possible occurrence of a seizure in the future. For example, the alarms can be utilized for therapeutic treatments to be administered to the patient for mitigating the pain \cite{elger2001future},\cite{dhulekar2015seizure}. Scalp EEG is one of the most prominent ways for predicting the onset of seizures \cite{esteller2005continuous}. The EEG of the epileptic patient can be classified into interictal period, preictal, ictal (during seizure) and post-ictal period. Detecting the transition from interictal to preictal period can help predict seizures in many cases of epilepsy, and there are many studies  which have shown the changes in the EEG of a patient when the patient transitions  from interictal to preictal periods. For example, Mirowski \emph{et al}. suggested that there is a decrease in the synchronization of the seizure focus region in the beta band range of the EEG. This is in contrast to an interictal period which involves moderate synchronization at higher frequency bands \cite{mirowski2009classification}.    

In this work, we investigate the possibility of using deep learning models for seizure prediction using single seizure preictal recording. Specifically, our approach uses Siamese network based similarity classifiers for predicting seizures. To the best of our knowledge, this is the first study to investigate the use of deep Siamese networks to multi-channel EEG data for seizure prediction. This is also the first study to investigate seizure prediction using single seizure recording. We also compare this approach with transfer learning of convolutional neural network (CNN) with one seizure and present the results for the same architecture with training using multiple seizures. This paper is organized as follows: In section \ref{sec:relatedwork}, we discuss the related work in the area of seizure prediction. In section \ref{sec:motivation}, we explain the benefit of using methods which can perform seizure prediction with lesser data for training. Section \ref{sec:methodology} explains the methodology of our proposed approach and the baselines utilized for comparison. Section \ref{sec:evaluation} shows the results of the proposed approach and compares it with two different baselines. Section \ref{sec:conclusion} summarizes the work and talks about future work. 

\section{Related Work}
\label{sec:relatedwork}

Seizure prediction requires the extraction of useful information from raw EEG signals. This information can be done manually or automatically (using CNN). One of the methods is to extract the features from the different EEG channels \cite{mormann2007seizure}. A rolling window of a predetermined size (e.g. 10 seconds) is utilized to extract various features for the model to be trained using supervised learning methods. Various features are extracted from the EEG channels. One of the first attempts to predict seizures investigated spectral properties of EEG data for different states of the patient with epilepsy \cite{Viglione}. It has been shown that multivariate channel features are better than univariate features like lyponov exponents \cite{DALESSANDRO2005506},\cite{Esteller},\cite{Harrison},\cite{iasemidis2005long}, \cite{jouny2005signal},\cite{le2005preictal},\cite{mormann2005predictability},\cite{dhulekar2015seizure}. Synchronization between channels is another important aspect that helps to differentiate between interictal and preictal states. Phase-locking between two different channels is useful for making this differentiation [37,38]. Progress made using multivariate signals has also led to studying the concepts from network theory to describe the graph based topology for the multivariate EEG time-series \cite{barrat2008dynamical},\cite{boccaletti2006complex},\cite{demir2005augmented},\cite{li2012effective},\cite{newman2003structure},\cite{strogatz2001exploring},\cite{dhulekar2015seizure}. After extracting these features, different methods like Markov processes \cite{lytton2008computer}, SVM \cite{chandaka2009cross},\cite{chisci2010real} and artificial neural networks \cite{liu2002multistage} have been utilized for performing seizure prediction.

Deep learning based algorithms are becoming more prominent for patient-specific prediction given the improved performance using various architectures. Every patient has different characteristics of the seizure because of which models are trained using the EEG from the specific patient. Daoud et al, used various deep learning architectures to show their performance of specific patients in CHB-MIT data \cite{daoud2019efficient}, \cite{shoeb2004epilepsy}. Among the most prominent is the use of convolution neural network (CNN) with a long-short term memory unit (LSTM) and a deep auto-encoder for performing seizure prediction. In doing so, the authors utilized the raw EEG signal for processing the incoming EEG stream. Khan \emph{et al}, utilized a CNN with wavelet transformed features to perform seizure prediction \cite{khan2017focal}. In \cite{tsiouris2018long}, the authors show results for LSTM architecture for various pre-seizure horizon windows for extracting the preictal features. These features range from various statistics like variance, skewness, kurtosis to spectral and graph theoretic features. These methods achieve significant improvement in prediction sensitivity with low false alarm rates. But the deep learning models inherently require a significant amount of data for training the models. This translates to using multiple seizures for a specific patient for learning a patient-specific model. As shown in \cite{kiral2018epileptic}, these models have to be periodically retrained because of non-stationary property of EEG. 

In this work, we investigate and compare the possibility of performing seizure prediction by training on a single seizure using deep models. The deep learning models require a lot of data. Some work has utilized transfer learning, but there is a need to investigate the performance for seizure prediction with very few seizures \cite{daoud2018deep},\cite{daoud2019efficient},\cite{raghu2020eeg},\cite{dhulekar2015seizure}. In this paper we utilize deep similarity based classification using Siamese networks for performing seizure prediction using one seizure data. Similarity based methods like similarity index \cite{le1999anticipating} have shown promise. We utilize CNN as the base network to Siamese network. This benefits by acting as an automatic features extraction algorithm from the wavelet transformed input data as utilized in \cite{khan2017focal}. The pairs of similar and dissimilar interictal-preictal examples increase the data and add to the variety of training sets.


\section{Motivation}
\label{sec:motivation}


Scalp EEG is a prominent method for predicting seizures. The problem requires the underlying systems to differentiate between the interictal and preictal time series EEG data. Recent advancements in  deep learning have helped in improving the seizure prediction accuracy using patient-specific models \cite{daoud2019efficient},\cite{tsiouris2018long}. Different patients have different interictal and preictal probability distributions of EEG. Hence patient-specific deep models perform better than generalized seizure prediction solutions, but these models require patient-specific interictal and preictal examples  \cite{daoud2019efficient}. Training deep models requires a significant amount of data, and the existing literature shows their utility by training using multiple seizures \cite{khan2017focal},\cite{tsiouris2018long},\cite{daoud2019efficient},
\cite{daoud2018deep}. For the purpose of training, the interictal data is usually abundant, but to gather the preictal examples for training, we have to wait for multiple seizures to occur. Although the lag between leading seizures depends upon the patient, it can be observed using CHB-MIT EEG data that this lag adds to the significant wait time apart from the energy, communication and storage requirements (most existing methods utilize EEG samples at 256 Hz) \cite{shoeb2004epilepsy}.Such high waiting times are not ideal from the stand point of bringing such systems into main stream usage, as it would be undesirable to the patient.

In case of an on-body device, there will be a need for transmitting high frequency data which will require a lot of energy for transmitting the data given that the device does not have deep learning hardware capabilities. At the same time, the need for recording more seizures requires more storage capabilities. Given the high frequency data, we can expect the data to be in gigabytes for each patient. Storing and communicating this data can also be problematic assuming that deep learning models will be trained at a local server. Furthermore, EEG is non-stationary, and it has been shown in \cite{kiral2018epileptic} that the learned model needs to be re-trained. Although re-tuning the weights of the models is possible, this process has to utilize more data; otherwise, the learned models will be prone to over-fitting. To overcome this problem, applications like face recognition employ few-shot learning based methods which utilize Siamese networks for learning a model which can provide a score for a similarity of two face images. We utilize a similar idea to help in solving the seizure prediction problem while at the same time utilizing less data. 

The Siamese learning model is able to develop combinations of similar and dissimilar examples for learning to classify the similarity of two given  test examples. The data set can be increased by posing the seizure prediction problem as a similarity/dissimilarity classification problem. This means that we can augment the data set size for training by taking just a few minutes of preictal EEG. In this work, we show our results by just utilizing 10 minutes of interictal and preictal with ictal EEG. Using the similarity learning method can provide several combinations of similar and dissimilar pairs of examples to add to the variety of data. This in turn helps in learning a generalized patient-specific model to classify two EEG examples as similar or dissimilar.
Our overall solution reduces the required data for training a deep learning model which is especially beneficial for solving some of the problems mentioned. Lower data requirements mean less communication, storage and energy usage. The retraining of the model due to the non-stationary property of EEG can be handled more effectively with lower data requirements. Hence, our solution augments the existing methods while mitigating problems for implementing practical seizure prediction systems.

\section{Methodology}
\label{sec:methodology}

\subsection{Data and Preprocessing} 
\label{sec:data_processing}
In this work, we utilize CHMIT scalp EEG data set \cite{shoeb2004epilepsy}. The dataset consists of scalp EEG recordings from 22 patient with variable amount of EEG recordings for each patient ranging from 9 hours to more than 50 hours. Each of the seizure is labeled, with a total of 136 seizures and 509 recordings of variable hours, belonging to an EEG without any seizures. The recordings were sampled at 256Hz. We selected a subset of these patients for our study with non-seizure recordings selected with similar assumptions as in \cite{daoud2019efficient} for testing the specificity of our method.      

Although the Siamese network helps us in differentiating the two incoming samples and to classify them as either similar or dissimilar, we still need the preictal examples to form the similar-dissimilar combinations for training the Siamese network. Algorithm \ref{alg:data_gathering} gives an overview of collecting the preictal and interictal data  from one of the seizures in a real-time practical scenario. It is assumed that the interictal recording is $m=4$ hours before and after the seizure onset. The overall idea is to collect the $t$ minutes of interictal samples $m$ hours before the seizure onset and similarly, $t$ minutes of preictal and ictal samples around the seizure onset. The value of $t$ used in our experiments is 10 minutes. This is a hyper parameter. The choice of $t$ is simply to make sure that the preictal data truly does belong to the preictal and ictal probability distribution since it could be possible that the preictal state of the patient begins just a few minutes before the seizure onset. This choice reduces the possibility of using the wrong examples for learning a classifier. The combination of samples helps in augmenting the data for Siamese learning which is otherwise not possible when utilizing a non-similarity based classifier or in case we perform fine-tuning of the already learnt model because of higher chances that the model might over fit to the data. 

The algorithm stores the $t$ minutes of the EEG samples denoted as $D_{I}$ and then waits for the seizure to occur at which point the system again stores another $t$ minutes of EEG samples $D_{P}$ which includes the preictal and ictal samples. Our assumption requires that the $D_{I}$ should be $m$ hours before the the seizure onset. In case the time before the seizure onset $m_{I}$, when $D_{I}$ was collected is less than $m$, then we record $D_{I}$ again until $m_{I}> m$.  It is possible that this process could go on for a long time, but in reality the fact that the system will have to wait for only one leading seizure to occur is highly likely. This is evident from the CHB-MIT data set where for most patients, we have a considerable gap between the leading seizures for data to be collected.

\begin{algorithm}[tb]
	\caption{An algorithm for real-time collection of preictal and interictal data. See section \ref{sec:data_processing} for explanation.}
	\label{alg:data_gathering}
	\begin{algorithmic}
		\WHILE{TRUE}
		\STATE $D_{I}$ $\leftarrow$ $t$ minutes of EEG recording
		\STATE $t_{1} \leftarrow $ GetCurrentTime()
		\WHILE{{\bfseries not} Seizure}
		\STATE wait till the seizure occurs
		\ENDWHILE
		\STATE $t_{2}$ = GetCurrentTime()
		\STATE $D_{p} \leftarrow$ $t$ minutes of EEG recording before seizure onset
		\IF{$t_{2}-t_{1}>m$}
		\STATE	\bfseries break
		\ENDIF
		\ENDWHILE
	\end{algorithmic}
\end{algorithm}

\subsubsection{Pre-processing}
\label{sec:Preprocessing}
After gathering the data, $D_{I}$ and $D_{p}$, our pre-processing step is similar to the one used in \cite{khan2017focal}. We use continuous wavelet transform to convert the incoming time series to get both the time and frequency information.  Taking the wavelet transform of the EEG signal converts it into time, frequency and channels mode. The time and frequency information helps in our analysis and due to this 3 dimensional structure of the tensor, we choose CNN which also serves as a feature extractor from the input. The CNN also acts as the base network for our Siamese network framework.
For the wavelet transform, we use the Mexican-hat mother wavelet with didactic scale from 1 to 512 as used in \cite{khan2017focal}. Each incoming sample is re-sampled to 128 HZ before taking the wavelet transform . The signal is then converted in 1 second window samples which correspond to 128 by 10 dimensional tensors for each of the channels. There is no overlap between the consecutive windows of 1 second. Each of these windows correspond to a separate example for the network during training and testing. Note that the Siamese network architecture will have two of these examples to serve as a similar-dissimilar example to a network as it is only learning to classify the similarity of the two examples. In case of traditional setup of classification using a CNN between preictal and interictal examples, this 1 second window tensor serves as an input. The CNN also serves as a baseline for comparing the results of our model trained using one seizure with model trained using multiple seizures. 

\subsubsection{Balancing and Preictal Length}
\label{sec:labelling}
For the seizure prediction problem, we have to deal with the class imbalance between the interictal and preictal examples. Possible methods to deal with this class imbalance include penalization of the majority class, under-sampling the majority class or oversampling the minority class. We under-sample the interictal class, and the algorithm \ref{alg:data_gathering} keeps this under consideration. As mentioned earlier, we utilize only the 10 minutes around the seizure onset for making up the preictal and ictal examples for training the model. This is a hyper parameter because the actual start of the preictal period is not known, and this period length may vary from patient to patient or from seizure to seizure. Our reasoning is that the underlying system can take the preictal state at any arbitrary time period. In order to ensure that the examples that we choose are from the preictal state of the patient, we try to get the examples as temporarily close to the seizure as possible. In the literature, different work investigates various preictal lengths. For example, in \cite{tsiouris2018long}, the authors utilize preictal lengths of 15, 30, 60 and 120 minutes for gathering the examples from the preictal class for training the LSTM model. It is not necessary that all the EEG for the pre-selected preictal length would consist of the EEG actually belonging to the preictal state of the patient. Hence, one will be confusing the classifier by providing examples which actually come from the interictal state even though they are from within the designated preictal length before the seizure onset. For models trained using multiple seizures, falsely labelled preictal samples might be compensated by multiple seizures. However for our solution, keeping the length of the preictal period to be as small as possible adds robustness and reduces the possibility of contaminating the training set with wrong labels. Although the raised alarm are based on a learnt models, it is highly likely that choosing the preictal length as close to the seizure will help in getting less contaminated preictal examples for training a model. This reduces preictal class size. In line with this reasoning, training using the traditional classification approach can cause problems like over-fitting if fine tuning/transfer learning is utilized due to lesser data. Siamese learning is also beneficial in this regard where we can use combinations of preictal-interictal class examples to augment the data set. Different examples also provide various combinations of similar-dissimilar examples for the Siamese network which help in generalization of the network. On the other hand, the traditional classification approach will not be able to generalize well given that we utilize less data. It should be noted that the preictal length is different from the prediction horizon. Prediction horizon is defined as the duration within which the seizure will occur given the prediction system raises an alarm. We have taken it to be different from the preictal length because the actual start of preictal period might vary from patient to patient or from seizure to seizure. For evaluating our method and the baselines, we have set the prediction horizon of 60 minutes.

\subsection{Seizure Prediction Using Siamese Learning}

In this subsection, we provide a generalized methodology for using the Siamese network for seizure prediction. Although we use wavelet transform as an input features to the network, the proposed methodology can easily be adapted for input of any other form. As mentioned earlier, few-shot learning using Siamese network has been useful for solving different problems like image recognition, face recognition etc. \cite{koch2015siamese}. Few-shot Siamese network is specially useful for cases where we have little data for training a model just like in face recognition. We adapt the same idea for seizure prediction using scalp EEG.  

\begin{figure*}[h]
	\includegraphics[width=1\textwidth,height = 0.3\textwidth]{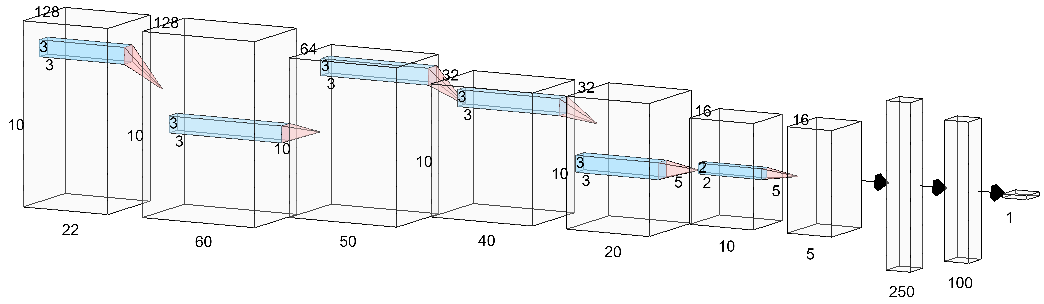}
	
	\caption{The CNN architecture used for performing non-similarity based seizure prediction. Each CNN layer is followed by a batch normalization layer with max pooling after every other layer. The parameters of this architecture are set based on the architecture in \cite{khan2017focal} since our problem formulation is very similar to this work.}
	\label{fig:CNNarchitecture}
\end{figure*}

The weights of the base network of the Siamese network are shared among each other. For seizure prediction, the inputs to the base network could either belong to the  preictal or interictal class. Since the Siamese network provides the similarity score of the two inputs, there will be finite possibilities for making similar and dissimilar class examples for the seizure prediction problem. As mentioned previously, if we are only utilizing preictal and interictal examples, then the possible combination for similar inputs can be interictal-interictal or preictal-preictal. The dissimilar class input to the network can only be preictal-interictal. Let $X_{a}$ and $X_{b}$ be the two inputs. The idea is to get the similarity score for the two inputs $X_{a}$ and $X_{b}$. Although for this work, the input to the network is the wavelet transformed EEG signal as described previously, for our methodology any transformed form of the input can be utilized. For our adaptation of Siamese networks we pose the problem as a similarity classifier. The score output, $S$, that quantifies the similarity between $X_{a}$ and $X_{b}$ can be written using equation \ref{eq:score}

\begin{equation}
	\label{eq:score}
	S = \frac{1}{1+e^{-h(X_{a},X_{b})}} 
\end{equation}

We use a sigmoid function for getting the output score which also scales the output between 0 and 1. In this case, a score close to 1 means that the two inputs are very similar. In equation \ref{eq:score}, the function $h(X_{a},X_{b})$ can be written as follows:

\begin{equation}
	\label{eq:network}
	h(X_{a},X_{b}) = g(|f(X_{a})-f(X_{b})|)
\end{equation}

In equation \ref{eq:network}, $f(\cdot)$ is any generalized base network for the Siamese network. For this work, we have utilized a CNN with a dense layer to represent $f(\cdot)$. It should be noted that any arbitrary network can be used for $f(\cdot)$. The absolute subtraction of the output of two base networks passes through the dense layer $g(\cdot)$ to be scaled by the sigmoid function as in equation \ref{eq:score}.

\subsubsection{Method A: Multiple Seizures}
\label{sec:architecture}

In order to show the significance of using our method, we train and show results for a traditional (non-similarity based) classifier which is trained using all the patient seizures to compare the results for the case when a single seizure is utilized versus the case when more than one seizure is utilized. Furthermore we also show the results by fine tuning the already trained model on the new patient's single seizure recording.

As described in section \ref{sec:relatedwork}, in recent approaches, the seizure prediction problem is solved using machine learning by training a deep learning model like CNN or recurrent neural network (RNN) for distinguishing between interictal and preictal states of the EEG. Our approach which uses multiple seizures utilizes a CNN with dense layer for classification between preictal and interictal samples. For easy reference, we call this method-A. We use the leave-one-out approach as in \cite{daoud2019efficient} for training Method-A. In this approach, out of the $N$ seizures, $N-1$ are used for training and the remaining 1 seizure is used to test the model. For training, 85\% of the data are used for the training set with 15\% for making the validation set. All the hyper-parameters are decided based on the validation set. Our network architecture largely follows the work in \cite{khan2017focal} since we are also utilizing similar pre-processing steps. 



For the Method-A and for good comparison, we also use the wavelet transformed pre-processed data as an input to the deep network model.  Figure \ref{fig:CNNarchitecture} shows the architecture used in this case which is motivated from the work in \cite{khan2017focal}. We use six convolutional layers with dense layer. The input to this network is a continuous time wavelet transform of 1 second 22 channel EEG signal. We perform batch normalization after every convolutional layer with drop out at the end of every layer. Max-pooling is performed after every other layer as shown in Figure \ref{fig:CNNarchitecture} with ReLU activation after every layer except for the final layer which uses sigmoid function. The model is trained using binary cross-entropy loss with Adam optimizer using Keras deep learning library in Python.

\vspace{0.1in}

\subsubsection{Siamese Network Architecture}
We have described the implementation of the CNN+NN classifier. This was needed for explaining the network architecture of the Siamese network approach mentioned earlier. We use only one seizure for getting the training set for learning the Siamese network model. Although we have given the algorithm \ref{alg:data_gathering} for getting the data in real time, for the purpose of experiments, we have multiple seizure data available per patient. For selected patients, we also verified the use of different patient-specific seizures during training for prediction, and the results indicate that different seizures results in similar prediction sensitivity. For our experiments, we select the seizure randomly and make training data which is a combination of similar and dissimilar class examples. Again we utilize 85\% of this data as a training set and rest as a validation set. This data set is used to train the Siamese network with base network $f(\cdot)$ as the CNN described for the method-A above (Only the architecture is similar, and the weights of CNN filters are trained from the collected one seizure data with no pre-training). This architecture constitutes the transformation $h(\cdot)$ in (\ref{eq:network}). This is followed by the absolute subtraction of the outputs of the CNN. The output is passed through two dense layers of 250 and 100 hidden units representing $g(\cdot)$. As before, RelU activation and dropout are added to these layers as well and sigmoid activation to give us the similarity score of the two examples under consideration. The model in this case is also trained using the Adam optimizer.

\subsubsection{Inference Stage}

After the model is trained, we perform  the inference  where the model gives us the classification for the incoming 1 second test EEG samples. The recordings used for testing are independent of the recordings used for training. We first describe the inference for the Siamese network method which uses one seizure data and then give a brief description for the inference for the CNN based model which uses multiple seizures during training.  

During training of the Siamese network, we have to make different combinations of similar and dissimilar classes using the preictal and interictal examples from $D_{I}$ and $D_{p}$. For testing we need to compare the incoming test sample with the predefined support set from the interictal and preictal class. For the Siamese network model, we utilize a few-shot based support set for comparing the similarity of the two incoming test samples with the interictal and preictal examples in the support set. The support set is made up of the randomly selected examples from $D_{I}$ and $D_{p}$ thus forming a set $S_{i}$ and $S_{p}$ for preictal and interictal examples respectively. For each of the incoming examples $X_{test}$, we generate the score using equation \ref{eq:score} corresponding to each of the test examples in $S_{I}$ and $S_{p}$. The score for the  support set examples is then averaged which forms the final score ($S_{p,t};S_{I,t}$) for sample $X_{test,t}$ belonging to the interictal or preictal class. The preictal and interictal score for the incoming samples are then smoothed and compared. If the resulting preictal score is greater than the interictal score above a certain threshold, then the corresponding test samples is given a score of 1 representing the decision by the model. The final decision is then made by looking at the window of consecutive decisions using an exponential averaging approach as used in \cite{khan2017focal}. The window size and the threshold for making the decision are hyper parameters in this case and are decided on the bases of the seizures collected earlier for a specific patient. The inference stage for the model trained using multiple seizures is similar to the one followed in \cite{khan2017focal}. The CNN is used for generating an output probability just like in \cite{khan2017focal}. The difference is that we pose the seizure prediction as binary classification where as Khan \emph{et al.} also separately includes the ictal class during training and combines the preictal and ictal probability during the inference stage using exponential averaging. We also introduce the exponential smoothing parameter which is inferred using the validation set.

\section{Evaluation}
\label{sec:evaluation}

\begin{table}[t]

	\begin{center}
		\begin{tabular}{cc|c|c|c|}

			\cline{3-5}
			\multicolumn{1}{l}{}                     & \multicolumn{1}{l|}{} & \multicolumn{3}{c|}{\textbf{Siamese Network/Method-A/Method-B}} \\ \hline
			\multicolumn{1}{|c|}{\textbf{ P }} & \textbf{ S }   & \textbf{\# Predicted}  & \textbf{Sensitivity}  & \textbf{fpr/h}  \\ \hline
			\multicolumn{1}{|c|}{1}                  & 7                     & 7/7/7                 & 100/100/100           & 0.08/0.08/0.08  \\ \hline
			\multicolumn{1}{|c|}{3}                  & 7                     & 7/7/6                 & 100/100/85.7          & 0.3/0.083/0.4 \\ \hline
			\multicolumn{1}{|c|}{6}                  & 10                     & 9/9/8                 & 90/90/80        & 0.07/0.083/0.07 \\ \hline
			\multicolumn{1}{|c|}{8}                  & 5                     & 5/5/5                 & 100/100/100           & 0/0/0           \\ \hline
			\multicolumn{1}{|c|}{10}                 & 7                     & 6/6/5                 & 85.7/85.7/71.4        & 0.12/0.08/0.2   \\ \hline
			\multicolumn{1}{|c|}{18}                 & 6                     & 5/5/4                 & 83.3/83.3/66.7         & 0.08/0.08/0.22  \\ \hline
			\multicolumn{1}{|c|}{20}                 & 7                     & 7/7/7                 & 100/100/100           & 0/0/0.03       \\ \hline
			\multicolumn{1}{l|}{}     & 49                    & 46/46/42              & 93.8/93.8/85.7        & 0.09/0.05/0.14  
			\\ \cline{2-5}
			
		\end{tabular}
		
	\end{center}
	\caption{Table shows the results for seizure prediction. Siamese network method uses a single seizure for predicting seizures. Method-A is the case when a leave-one-out approach is used for training the model i.e. N-1 out of N seizure used for training a CNN classifier. Method-B is the same as method-A except that only 1 seizure with 10 minutes of preictal data is utilized for fine tuning the model learned on another patient. (\textbf{P denotes Patient \#, S denotes \# Seizures, fpr/h denotes false predictions per hour})
	}
	\label{tab:results}
\end{table}

\begin{figure} 
	\centering
	\includegraphics[width=1\linewidth]{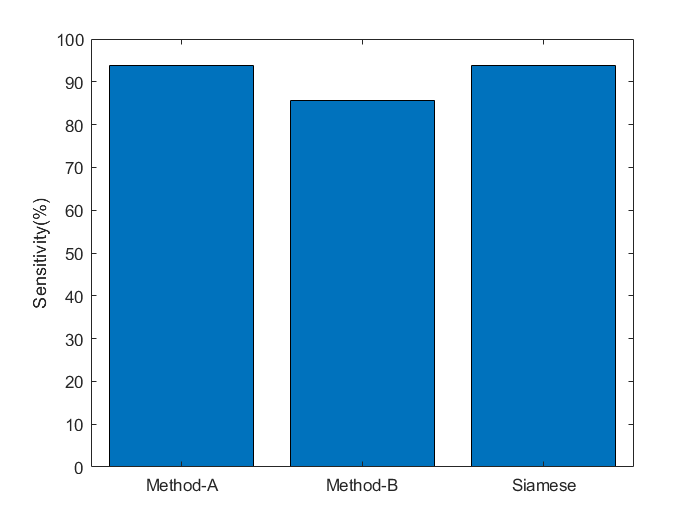}
	\caption{\footnotesize Comparison of the overall seizure sensitivity for the three methods.The overall sensitivity for Siamese learning is better than method-B} 
	\label{fig:barplot}
\end{figure}

In this section, we describe the results for the Siamese network using one seizure. We also show the results for the case when multiple seizures are used for training the CNN model as described in section \ref{sec:architecture} (Method-A). As previously mentioned, a leave-one-out approach is utilized for evaluating method-A in which $N-1$ seizures are used for training the model for testing each of the $N$ seizures. Method B uses the same underlying method as method-A except that only 1 seizure is utilized for getting training examples for training the model. This training set is used for transfer learning on a model already trained using multiple seizures from another patient. 

A seizure is predicted correctly if our method raises an alarm and the seizure occurs within the prediction horizon. The prediction horizon for all the methods in this work is taken to be one hour. Although the duration of the prediction horizon is one hour, the time before the seizure from which we choose our preictal examples can vary depending upon the method utilized. The reason for choosing a different length is argued in section \ref{sec:labelling}. For the method-A, we choose the preictal length to be 15 minutes from every seizure and 10 minutes for method-B from one seizure only for standard comparison with Siamese networks. As pointed out earlier, there are few alternatives other than using our Siamese network model. We can train method-A with just one seizure without any transfer learning from another patient model. In this case, the trained model does not performs well, and most of the seizures are not predicted. Another alternative is to use transfer learning (method-B) using an already trained model and fine tuning using the data from the patient under consideration. We show the results in this case. During the training phase for the alternative method (B), we found that the trained model was not robust and was dependent upon the underlying patients weights used for transfer learning. We evaluate these methods by finding the sensitivity of predicting seizures and false predictions per hour for each patient. A raised alarm is a false prediction given that the method raises an alarm for a seizure and the seizure does not occur within the prediction horizon. 

Table \ref{tab:results} and figure \ref{fig:barplot} show the comparison of three different methods on 306 hours of scalp EEG recordings with a total of 49 seizures. Our Siamese network method, which uses one seizure for training, achieved an overall sensitivity of 93.8\% which is the same as Method-A in which we utilized multiple seizures for training. Method-B which utilizes transfer learning using one seizure achieved 85.7\% seizures. Although the best model in this case is still the Method-A, Siamese learning is able to achieve the same sensitivity and an overall false predictions rate of 0.09/hr using single seizure data. This results in 2.4 false predictions in 24 hours compared to 1.2 for method-A. The transfer learning approach (method-B) has a relatively lower sensitivity of 85.7\% with 0.14 fpr/h (false positives). It can be observed from Table \ref{tab:results}, that for some patients, all three methods have comparable results, with method-B lagging in a few cases. As pointed out earlier, we also trained the architecture for method-A without transfer learning for different patients. The results showed very low sensitivity with many false predictions. This shows that having pre-trained weights from another patient model helps with the training process under less data. It should be noted that we are using a preictal time for alarming the patient. Different methods in the literature report various measures of accuracy like specificity and sensitivity for varying prediction time (training data for preictal period is also selected from this horizon) which makes the comparison with these existing methods difficult. It is for this very reason, we have utilized the given method-A and method-B as baseline to show the utility of the proposed method while still utilizing lesser seizures during training. Keeping this under consideration, we would direct the reader to table 3 and 4 in \cite{daoud2019efficient} for comparison of our work with the existing solutions. It can be seen that our method achieves comparable sensitivity and false alarm rate while at the same time utilizing far less seizure data. 

From the perspective of online learning for seizure prediction, our study shows that training and fine tuning using Siamese learning will be a useful addition to existing seizure prediction methods. Although we have designed the base-network of the Siamese network keeping with the baselines and architecture from \cite{khan2017focal}, different base line architectures will surely help with improving the performance of the system and is left for future work. It should be noted that we have shown the results for training on a single seizure. The approach can be extended to online learning without processing the bulk of the EEG signal with online learning with multiple seizures. It is also possible that the patient for whom the model is trained using a single seizure might have multiple variations of seizure types and the proposed Siamese network approach will help to generalize. This is the drawback for training with one seizure, but the similarity based approach helps in mitigating this possibility which can be seen from the sensitivity of Siamese learning to seizure prediction as compared to that of the transfer learning approach (method-B). Furthermore, extending this method to online learning will further help in generalizing our method.

\section{Conclusion and Future Work}
\label{sec:conclusion}
In this paper, we investigate the performance of deep learning methods for seizure prediction under the constraint of learning from just one seizure which will help towards easy adaptation of practical seizure prediction systems. We propose a similarity based Siamese network classifier approach with CNN as the base network for predicting seizures. We use the CHB-MIT scalp EEG data set to train and evaluate patient-specific seizure prediction models. As a baseline, we show the results for the the traditional CNN based classifier trained using multiple patient-specific seizures and with transfer learning with one seizure for a specific patient. The results show that our Siamese network approach, which utilizes one seizure for training, is able to achieve similar performance to a method that utilizes multiple seizures. The low data requirements of our method will make it easier to use off-the-shelf seizure prediction devices, adding to the utility for patients. In future work, we plan to convert this approach to an online learning system with the addition of training online on multiple seizures. In the literature, different deep learning architectures like LSTM with CNN have been shown to improve the performance of seizure prediction. A study on improving the proposed approach by testing with different base networks will also be useful for improving the performance of the system.

\bibliographystyle{IEEEtran}
\bibliography{example_paper}

\begin{thebibliography}{10}
\providecommand{\url}[1]{#1}
\csname url@samestyle\endcsname
\providecommand{\newblock}{\relax}
\providecommand{\bibinfo}[2]{#2}
\providecommand{\BIBentrySTDinterwordspacing}{\spaceskip=0pt\relax}
\providecommand{\BIBentryALTinterwordstretchfactor}{4}
\providecommand{\BIBentryALTinterwordspacing}{\spaceskip=\fontdimen2\font plus
\BIBentryALTinterwordstretchfactor\fontdimen3\font minus
  \fontdimen4\font\relax}
\providecommand{\BIBforeignlanguage}[2]{{%
\expandafter\ifx\csname l@#1\endcsname\relax
\typeout{** WARNING: IEEEtran.bst: No hyphenation pattern has been}%
\typeout{** loaded for the language `#1'. Using the pattern for}%
\typeout{** the default language instead.}%
\else
\language=\csname l@#1\endcsname
\fi
#2}}
\providecommand{\BIBdecl}{\relax}
\BIBdecl

\bibitem{fisher2005epileptic}
R.~S. Fisher, W.~V.~E. Boas, W.~Blume, C.~Elger, P.~Genton, P.~Lee, and
  J.~Engel~Jr, ``Epileptic seizures and epilepsy: definitions proposed by the
  international league against epilepsy (ilae) and the international bureau for
  epilepsy (ibe),'' \emph{Epilepsia}, vol.~46, no.~4, pp. 470--472, 2005.

\bibitem{valentinuzzi2007bioelectrical}
M.~E. Valentinuzzi, ``Bioelectrical signal processing in cardiac and
  neurological applications and electromyography: physiology, engineering, and
  noninvasive applications,'' \emph{BioMedical Engineering OnLine}, vol.~6,
  p.~27, 2007.

\bibitem{blume2001glossary}
W.~T. Blume, H.~O. L{\"u}ders, E.~Mizrahi, C.~Tassinari, W.~van Emde~Boas, and
  J.~Engel~Jr, Ex-officio, ``Glossary of descriptive terminology for ictal
  semiology: report of the ilae task force on classification and terminology,''
  \emph{Epilepsia}, vol.~42, no.~9, pp. 1212--1218, 2001.

\bibitem{van2014functional}
P.~Van~Mierlo, M.~Papadopoulou, E.~Carrette, P.~Boon, S.~Vandenberghe,
  K.~Vonck, and D.~Marinazzo, ``Functional brain connectivity from eeg in
  epilepsy: Seizure prediction and epileptogenic focus localization,''
  \emph{Progress in neurobiology}, vol. 121, pp. 19--35, 2014.

\bibitem{daoud2019efficient}
H.~{Daoud} and M.~A. {Bayoumi}, ``Efficient epileptic seizure prediction based
  on deep learning,'' \emph{IEEE Transactions on Biomedical Circuits and
  Systems}, vol.~13, no.~5, pp. 804--813, 2019.

\bibitem{brodie2002staged}
M.~J. Brodie and P.~Kwan, ``Staged approach to epilepsy management,''
  \emph{Neurology}, vol.~58, no. 8 suppl 5, pp. S2--S8, 2002.

\bibitem{elger2001future}
C.~E. Elger, ``Future trends in epileptology,'' \emph{Current Opinion in
  Neurology}, vol.~14, no.~2, pp. 185--186, 2001.

\bibitem{dhulekar2015seizure}
N.~Dhulekar, S.~Nambirajan, B.~Oztan, and B.~Yener, ``Seizure prediction by
  graph mining, transfer learning, and transformation learning,'' in
  \emph{International Workshop on Machine Learning and Data Mining in Pattern
  Recognition}.\hskip 1em plus 0.5em minus 0.4em\relax Springer, 2015, pp.
  32--52.

\bibitem{esteller2005continuous}
R.~Esteller, J.~Echauz, M.~D'Alessandro, G.~Worrell, S.~Cranstoun,
  G.~Vachtsevanos, and B.~Litt, ``Continuous energy variation during the
  seizure cycle: towards an on-line accumulated energy,'' \emph{Clinical
  neurophysiology}, vol. 116, no.~3, pp. 517--526, 2005.

\bibitem{mirowski2009classification}
P.~Mirowski, D.~Madhavan, Y.~LeCun, and R.~Kuzniecky, ``Classification of
  patterns of eeg synchronization for seizure prediction,'' \emph{Clinical
  neurophysiology}, vol. 120, no.~11, pp. 1927--1940, 2009.

\bibitem{mormann2007seizure}
F.~Mormann, R.~G. Andrzejak, C.~E. Elger, and K.~Lehnertz, ``Seizure
  prediction: the long and winding road,'' \emph{Brain}, vol. 130, no.~2, pp.
  314--333, 2007.

\bibitem{Viglione}
\BIBentryALTinterwordspacing
S.~Viglione and G.~Walsh, ``Proceedings: Epileptic seizure prediction,''
  \emph{Electroencephalography and clinical neurophysiology}, vol.~39, no.~4,
  p. 435—436, October 1975. [Online]. Available:
  \url{http://europepmc.org/abstract/MED/51767}
\BIBentrySTDinterwordspacing

\bibitem{DALESSANDRO2005506}
\BIBentryALTinterwordspacing
M.~D'Alessandro, G.~Vachtsevanos, R.~Esteller, J.~Echauz, S.~Cranstoun,
  G.~Worrell, L.~Parish, and B.~Litt, ``A multi-feature and multi-channel
  univariate selection process for seizure prediction,'' \emph{Clinical
  Neurophysiology}, vol. 116, no.~3, pp. 506 -- 516, 2005. [Online]. Available:
  \url{http://www.sciencedirect.com/science/article/pii/S1388245704004560}
\BIBentrySTDinterwordspacing

\bibitem{Esteller}
R.~Esteller, J.~Echauz, M.~D'Alessandro, G.~Worrell, S.~Cranstoun,
  G.~Vachtsevanos, and B.~Litt, ``Continuous energy variation during the
  seizure cycle: Towards an on-line accumulated energy,'' \emph{Clinical
  neurophysiology : official journal of the International Federation of
  Clinical Neurophysiology}, vol. 116, pp. 517--26, 04 2005.

\bibitem{Harrison}
\BIBentryALTinterwordspacing
M.~A.~F. Harrison, M.~G. Frei, and I.~Osorio, ``Accumulated energy revisited,''
  \emph{Clinical neurophysiology : official journal of the International
  Federation of Clinical Neurophysiology}, vol. 116, no.~3, p. 527—531, March
  2005. [Online]. Available: \url{https://doi.org/10.1016/j.clinph.2004.08.022}
\BIBentrySTDinterwordspacing

\bibitem{iasemidis2005long}
L.~Iasemidis, D.-S. Shiau, P.~M. Pardalos, W.~Chaovalitwongse, K.~Narayanan,
  A.~Prasad, K.~Tsakalis, P.~R. Carney, and J.~C. Sackellares, ``Long-term
  prospective on-line real-time seizure prediction,'' \emph{Clinical
  Neurophysiology}, vol. 116, no.~3, pp. 532--544, 2005.

\bibitem{jouny2005signal}
C.~C. Jouny, P.~J. Franaszczuk, and G.~K. Bergey, ``Signal complexity and
  synchrony of epileptic seizures: is there an identifiable preictal period?''
  \emph{Clinical neurophysiology}, vol. 116, no.~3, pp. 552--558, 2005.

\bibitem{le2005preictal}
M.~Le~Van~Quyen, J.~Soss, V.~Navarro, R.~Robertson, M.~Chavez, M.~Baulac, and
  J.~Martinerie, ``Preictal state identification by synchronization changes in
  long-term intracranial eeg recordings,'' \emph{Clinical Neurophysiology},
  vol. 116, no.~3, pp. 559--568, 2005.

\bibitem{mormann2005predictability}
F.~Mormann, T.~Kreuz, C.~Rieke, R.~G. Andrzejak, A.~Kraskov, P.~David, C.~E.
  Elger, and K.~Lehnertz, ``On the predictability of epileptic seizures,''
  \emph{Clinical neurophysiology}, vol. 116, no.~3, pp. 569--587, 2005.

\bibitem{barrat2008dynamical}
A.~Barrat, M.~Barthelemy, and A.~Vespignani, \emph{Dynamical processes on
  complex networks}.\hskip 1em plus 0.5em minus 0.4em\relax Cambridge
  university press, 2008.

\bibitem{boccaletti2006complex}
S.~Boccaletti, V.~Latora, Y.~Moreno, M.~Chavez, and D.~Hwang, ``Complex
  networks: Structure and dynamics physics reports, vol. 424,'' 2006.

\bibitem{demir2005augmented}
C.~Demir, S.~H. Gultekin, and B.~Yener, ``Augmented cell-graphs for automated
  cancer diagnosis,'' \emph{Bioinformatics}, vol.~21, no. suppl\_2, pp.
  ii7--ii12, 2005.

\bibitem{li2012effective}
G.~Li, M.~Semerci, B.~Yener, and M.~J. Zaki, ``Effective graph classification
  based on topological and label attributes,'' \emph{Statistical Analysis and
  Data Mining: The ASA Data Science Journal}, vol.~5, no.~4, pp. 265--283,
  2012.

\bibitem{newman2003structure}
M.~E. Newman, ``The structure and function of complex networks,'' \emph{SIAM
  review}, vol.~45, no.~2, pp. 167--256, 2003.

\bibitem{strogatz2001exploring}
S.~H. Strogatz, ``Exploring complex networks,'' \emph{nature}, vol. 410, no.
  6825, pp. 268--276, 2001.

\bibitem{lytton2008computer}
W.~W. Lytton, ``Computer modelling of epilepsy,'' \emph{Nature Reviews
  Neuroscience}, vol.~9, no.~8, pp. 626--637, 2008.

\bibitem{chandaka2009cross}
S.~Chandaka, A.~Chatterjee, and S.~Munshi, ``Cross-correlation aided support
  vector machine classifier for classification of eeg signals,'' \emph{Expert
  Systems with Applications}, vol.~36, no.~2, pp. 1329--1336, 2009.

\bibitem{chisci2010real}
L.~Chisci, A.~Mavino, G.~Perferi, M.~Sciandrone, C.~Anile, G.~Colicchio, and
  F.~Fuggetta, ``Real-time epileptic seizure prediction using ar models and
  support vector machines,'' \emph{IEEE Transactions on Biomedical
  Engineering}, vol.~57, no.~5, pp. 1124--1132, 2010.

\bibitem{liu2002multistage}
H.~S. Liu, T.~Zhang, and F.~S. Yang, ``A multistage, multimethod approach for
  automatic detection and classification of epileptiform eeg,'' \emph{IEEE
  Transactions on biomedical engineering}, vol.~49, no.~12, pp. 1557--1566,
  2002.

\bibitem{shoeb2004epilepsy}
A.~Shoeb, H.~Edwards, J.~Connolly, B.~Bourgeois, S.~T. Treves, J.~Guttag, and
  P.-S. S.~O. Detection, ``Epilepsy and behavior,'' \emph{August}, vol.~5,
  no.~4, pp. 483--498, 2004.

\bibitem{khan2017focal}
H.~Khan, L.~Marcuse, M.~Fields, K.~Swann, and B.~Yener, ``Focal onset seizure
  prediction using convolutional networks,'' \emph{IEEE Transactions on
  Biomedical Engineering}, vol.~65, no.~9, pp. 2109--2118, 2017.

\bibitem{tsiouris2018long}
K.~M. Tsiouris, V.~C. Pezoulas, M.~Zervakis, S.~Konitsiotis, D.~D. Koutsouris,
  and D.~I. Fotiadis, ``A long short-term memory deep learning network for the
  prediction of epileptic seizures using eeg signals,'' \emph{Computers in
  biology and medicine}, vol.~99, pp. 24--37, 2018.

\bibitem{kiral2018epileptic}
I.~Kiral-Kornek, S.~Roy, E.~Nurse, B.~Mashford, P.~Karoly, T.~Carroll,
  D.~Payne, S.~Saha, S.~Baldassano, T.~O'Brien \emph{et~al.}, ``Epileptic
  seizure prediction using big data and deep learning: toward a mobile
  system,'' \emph{EBioMedicine}, vol.~27, pp. 103--111, 2018.

\bibitem{daoud2018deep}
H.~Daoud and M.~Bayoumi, ``Deep learning based reliable early epileptic seizure
  predictor,'' in \emph{2018 IEEE Biomedical Circuits and Systems Conference
  (BioCAS)}.\hskip 1em plus 0.5em minus 0.4em\relax IEEE, 2018, pp. 1--4.

\bibitem{raghu2020eeg}
S.~Raghu, N.~Sriraam, Y.~Temel, S.~V. Rao, and P.~L. Kubben, ``Eeg based
  multi-class seizure type classification using convolutional neural network
  and transfer learning,'' \emph{Neural Networks}, vol. 124, pp. 202--212,
  2020.

\bibitem{le1999anticipating}
M.~Le~Van~Quyen, J.~Martinerie, M.~Baulac, and F.~Varela, ``Anticipating
  epileptic seizures in real time by a non-linear analysis of similarity
  between eeg recordings,'' \emph{Neuroreport}, vol.~10, no.~10, pp.
  2149--2155, 1999.

\bibitem{koch2015siamese}
G.~Koch, R.~Zemel, and R.~Salakhutdinov, ``Siamese neural networks for one-shot
  image recognition,'' in \emph{ICML deep learning workshop}, vol.~2.\hskip 1em
  plus 0.5em minus 0.4em\relax Lille, 2015.

\end{thebibliography}

\end{document}